\documentclass{article}

\usepackage{PRIMEarxiv}

\usepackage[utf8]{inputenc} 
\usepackage[T1]{fontenc}    
\usepackage{url}            
\usepackage{booktabs}       
\usepackage{amsfonts}       
\usepackage{nicefrac}       
\usepackage{microtype}      
\usepackage{lipsum}
\usepackage{fancyhdr}       
\usepackage{graphicx}       
\graphicspath{{media/}}     

\usepackage{multirow}%
\usepackage{amsmath,amssymb,amsfonts}%
\usepackage{amsthm}%
\usepackage{mathrsfs}%
\usepackage[title]{appendix}%
\usepackage{xcolor}%
\usepackage{textcomp}%
\usepackage{manyfoot}%
\usepackage{booktabs}%
\usepackage{algorithm}%
\usepackage{algorithmicx}%
\usepackage{algpseudocode}%
\usepackage{listings}%
\newtheorem{definition}{Definition}%

\title{Graph burning: an overview of mathematical programs
}

\author{
  \textbf{Lourdes Beatriz Cajica-Maceda, Freddy Alejandro Chaurra-Gutiérrez,} \\ \textbf{Julio César Pérez-Sansalvador, Jesús García-Díaz} \\
  Coordinación de Ciencias Computacionales \\
  Instituto Nacional de Astrofísica, Óptica y Electrónica \\
  Puebla, Mexico\\
  \texttt{\{bcajica,chaura,jcp.sansalvador,jesgadiaz\}@inaoep.mx} \\
}

\begin{document}
\maketitle

\begin{abstract}
The Graph Burning Problem (GBP) is a combinatorial optimization problem that has gained relevance as a tool for quantifying a graph's vulnerability to contagion. Although it is based on a very simple propagation model, its decision version is NP-complete, and its optimization version is NP-hard. Many of its theoretical properties across different graph families have been thoroughly explored, and numerous interesting variants have been proposed. This paper reports novel mathematical programs for the optimization version of the classical GBP. Among the presented programs are a Mixed-Integer Linear Program (MILP), a Constraint Satisfaction Problem (CSP), two Integer Linear Programs (ILP), and two Quadratic Unconstrained Binary Optimization (QUBO) problems. Most optimization solvers can handle these, being QUBO problems of a capital interest in quantum computing. The primary aim of this paper is to gain a comprehensive understanding of the GBP by examining its different formulations. Compared to other mathematical programs from the literature, the ones presented here are conceptually simpler and involve fewer variables. These make them more practical for finding optimal solutions using optimization algorithms and solvers, as we show by solving some instances with millions of vertices in just a few minutes.
\end{abstract}


\section{Introduction}\label{sec1}

Propagation phenomena are of significant interest to humanity. For example, by modeling the spread of fires, containment strategies can be developed. The same idea applies to a wide range of processes, including the diffusion of social influence, viruses, rumors, alerts, and news. Since the underlying structure of complex systems is a network, assessing a network's susceptibility to contagion is crucial. The Graph Burning Problem (GBP) provides a framework for addressing this need.

The optimization version of the GBP receives a graph $G$ as input and seeks a minimum-length burning sequence. Namely, a list $(u_1,u_2,...,u_g) \in V(G)^g$ with minimum $g$ and such that all vertices are at most at distance $g-i$ to some vertex $u_i$. The length of the optimal burning sequence is known as the burning number and is denoted by $b(G)$ \cite{bonato2014burning,alon1992transmitting}. This number captures how susceptible to contagion a graph is; the smaller the burning number, the more vulnerable the graph. The decision version of the GBP receives a graph $G=(V,E)$ and a positive integer $g$ as input. Its goal is to determine whether $b(G) \leq g$. This problem is NP-complete, even for trees of maximum degree three, spider graphs, and path-forests, among others \cite{bonato2020survey}. 

\begin{figure}[h]
\centering
\includegraphics[width=1\textwidth]{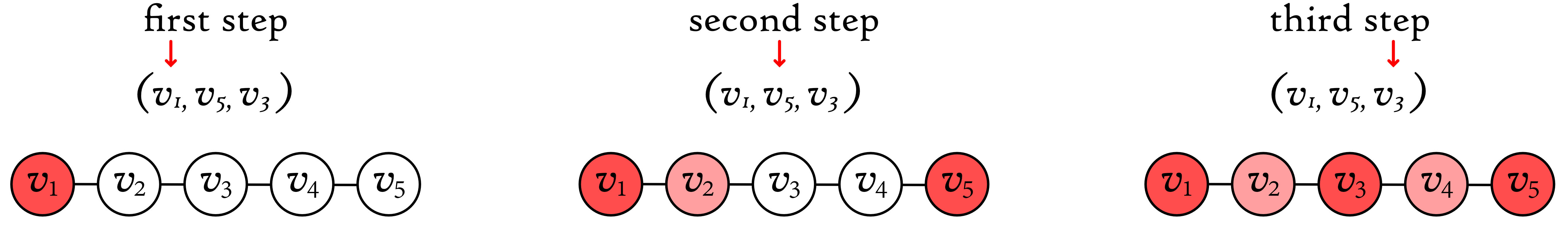}
\caption{The propagation process of the optimal burning sequence $(v_1,v_5,v_3)$ for the graph $P_5$. At every discrete step, vertices get \textit{burned} if they are in the neighborhood of a previously \textit{burned} vertex or if they are in the burning sequence at the current step. Notice that $b(P_5)=3$}\label{fig:1}
\end{figure}

\begin{figure}[h]
\centering
\includegraphics[width=1\textwidth]{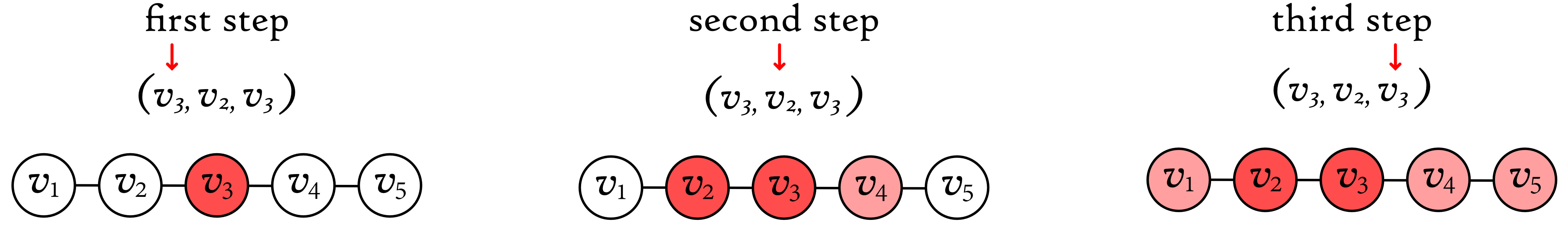}
\caption{The propagation process of the optimal burning sequence $(v_3,v_2,v_3)$ for the graph $P_5$. Notice that a burning sequence may have repeated vertices and a vertex might have multiple \textit{fire sources}. For instance, $v_3$ catches \textit{fire} from $v_3$ and $v_2$}\label{fig:2}
\end{figure}

Figures \ref{fig:1} and \ref{fig:2} show how the GBP is based on a simple discrete propagation process where \textit{fire} spreads to neighboring vertices. Initially, all vertices are \textit{unburned} and once a vertex is \textit{burned}, it remains in that state. Notice that a vertex can be \textit{burned} by many \textit{fire sources}, i.e., it can be at distance $g-i$ from many vertices in the burning sequence; this will be relevant later when introducing Quadratic Unconstrained Binary Optimization (QUBO) problems.

This paper introduces novel mathematical programs for the GBP. The goal of these is to have a better understanding of the problem. In addition, thanks to optimization algorithms and solvers, these programs let us search for optimal or near-optimal solutions. Some of the programs introduced are very similar to the previously reported ones (see Sections \ref{sec:ilp1} and \ref{sec:ilp2}); however, most of them are completely different. Before continuing, let us list some definitions and relationships that will be useful for the remainder of the paper.

\section{Main definitions and relationships}

\begin{definition}
    A graph $G = (V, E)$ is an ordered pair, where $V$ is the set of vertices and $E$ is the set of edges, a set of 2-element subsets of $V$ \cite{diestel2024graph}.\\
\end{definition}

\begin{definition}
    The distance $d(u, v)$ between the vertices $u$ and $v$ on a graph is the length of their shortest path.\\
\end{definition}

\begin{definition}
    The open neighborhood $N(v)$ of a vertex $v$ is its set of adjacent vertices in $G$.\\
\end{definition}

\begin{definition}
    The closed neighborhood $N[v]$ of a vertex $v$ is $N(v) \cup \{v\}$.\\
\end{definition}

\begin{definition}
    The $r^{th}$ power $G^r$ of a given graph $G = (V, E)$ is the result of adding an edge between each pair of vertices in $V$ at a distance of up to $r$.\\
\end{definition}

\begin{definition}
    The $r^{th}$ open neighborhood $N_r(v)$ of a vertex $v$ is its set of neighbors in $G^r$.\\
\end{definition}

\begin{definition}
    The $r^{th}$ closed neighborhood $N_r[v]$ of a vertex $v$ is $N_r(v) \cup \{v\}$.\\
\end{definition}

\begin{definition}
    A burning sequence of a graph $G = (V, E)$ is an ordered list $(u_1, u_2, ..., u_g) \in V^g$ such that the distance from every $v \in V$ to some vertex $u_i$ is at most $g - i$. The length of the burning sequence is $g$.\\
\end{definition}

\begin{definition}
    Given an input graph, the GBP seeks a burning sequence of minimum length.\\ 
\end{definition}

\begin{definition}
    The length of an optimal solution for the GBP is denoted by $b(G)$, which is known as the burning number of the graph.
\end{definition}

\subsection{The GBP as a coverage problem}






The GBP can be stated as finding the shortest sequence $(u_1,u_2,...,u_{g})$ that satisfies Eq. (\ref{eq:3}) \cite{bonato2020survey}.

\begin{equation}
    N_{g-1}[u_1] \cup N_{g-2}[u_2] \cup \cdots \cup N_{1}[u_{g-1}]\cup N_0[u_{g}] = V(G) \ \ .
    \label{eq:3}
\end{equation}

\begin{figure}[h]
\centering
\includegraphics[width=0.6\textwidth]{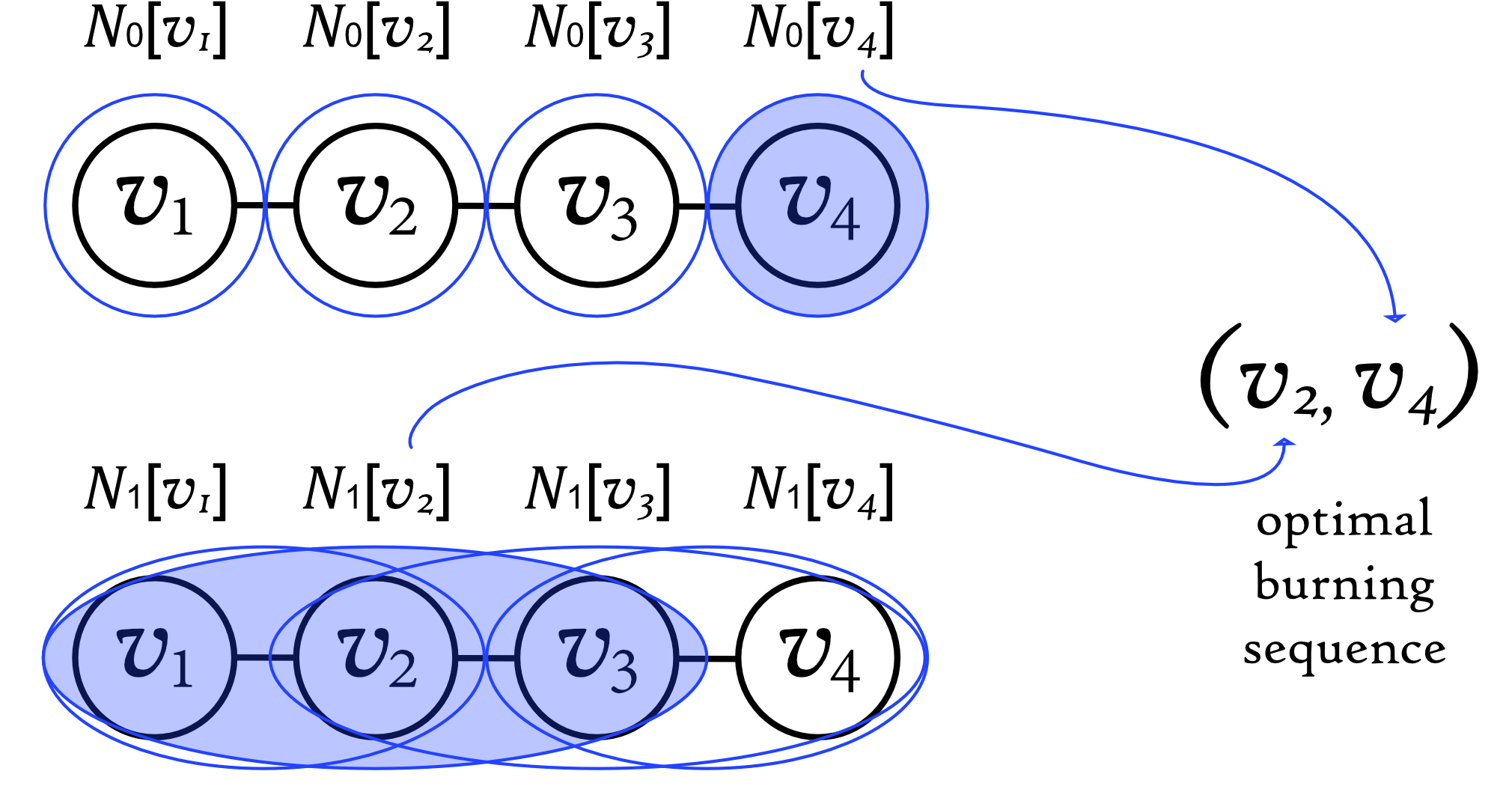}
\caption{The CMCP corresponding to $P_4$ with $g=b(P_4)=2$}\label{fig:3}
\end{figure}

From this equation, it is inferred that the GBP can be reduced to a series of NP-hard Clustered Maximum Coverage Problems (CMCPs). In summary, for a given guess $g\ge b(G)$, Eq. (\ref{eq:3}) can be satisfied by a sequence $(u_1,u_2,...,u_g)$ such that each vertex $u_j$ corresponds to a \textit{cluster} $\{ N_{g-j} [v] : v \in V(G) \}$. For example, if the input graph is $P_4$ and $g=b(P_4)=2$, its corresponding CMCP is shown in Figure \ref{fig:3}. As illustrated in this figure, a vertex must be chosen for each available \textit{covering radius}; 0 and 1 in this case. In this way, vertex $v_4$ is selected from the cluster $\{N_0[v_1],N_0[v_2],N_0[v_3],N_0[v_4]\}$ and vertex $v_2$ is selected from the cluster $\{N_1[v_1],N_1[v_2],N_1[v_3],N_1[v_4]\}$. Notice that $N_0[v_4] \cup N_1[v_2] = V(P_4)$. Thus, an optimal burning sequence for $P_4$ is $(v_2,v_4)$.

In case $g<b(G)$, Eq. (\ref{eq:3}) cannot be satisfied. Thus, a binary search can be performed to find an optimal solution (see Algorithm \ref{alg:1}). In short, the GBP can be solved by solving up to $\log_2 n$ CMCPs, where $n=|V(G)|$ and $g$ is a guess on $b(G)$ (see line 3 of Algorithm \ref{alg:1}). If the optimal solution of CMCP$(G,g)$ covers all vertices, the upper bound of the binary search is set to $g-1$ (see line 6 of Algorithm \ref{alg:1}); otherwise, the lower bound is set to $g+1$ (see line 8 of Algorithm \ref{alg:1}). In this way, in some iteration of the binary search, $g$ equals $b(G)$. Therefore, the GBP is solved to optimality. Most of the mathematical programs introduced in this paper leverage the relationship between the GBP and the CMCP.

For more details on the CMCP, the reader can refer to~\cite{garcia2025greedy} and~\cite{chekuri2004maximum}. In fact, when García-Díaz et al. proposed the CMCP in 2025~\cite{garcia2025greedy} they were not aware of this being a special case of the Maximum Coverage Problem with Group
Budget Constraints introduced by Chekuri and Kumar in 2004~\cite{chekuri2004maximum}.

\begin{algorithm}
\caption{GBP as a series of CMCPs}
    \begin{algorithmic}[1]
        \Require A graph $G=(V,E)$
        \Ensure An optimal burning sequence
        \State $l,u \Leftarrow 1,|V|$
        \While{$l \le u$}
        \State $g \Leftarrow \lfloor (l+u)/2 \rfloor$
        \State $S \Leftarrow \text{Solve CMCP}(G,g)$
        \If{$S$ burns all vertices}
        \State $u \Leftarrow g-1$
        \Else 
        \State $l \Leftarrow g+1$ 
        \EndIf
        \EndWhile \\
        \Return the best found burning sequence
    \end{algorithmic}
    \label{alg:1}
\end{algorithm}

\subsection{Mathematical programs}

Mathematical programs consist of three main components: an objective function, a set of constraints, and decision variables. Depending on the characteristics of these, a mathematical program belongs to a different category. For instance, a Linear Program (LP) has a linear objective function, linear constraints, and continuous variables:

\begin{align}
	\textbf{min} & \ \ \textbf{c}\textbf{x} &  \label{LP:1} \\
	\textbf{subject to (s.t.)}
	& \ \ \textbf{Ax} \le \textbf{b} &  \label{LP:2}\\
    \textbf{where}
	& \ \ \textbf{x} \ge 0 &   \label{LP:3}
\end{align}

The row vector $\textbf{c}$, the matrix $\textbf{A}$, and the column vector $\textbf{b}$ are the input data. The column vector $\textbf{x}$ contains the variables to optimize. If all of these variables are integers, then the program is an Integer Linear Program (ILP). If some variables are integer and some are continuous, then the program is a Mixed-Integer Linear Program (MILP). The mathematical programs listed in this paper include one MILP (see Section \ref{sec:milp}), two ILPs (See Sections \ref{sec:ilp1} and \ref{sec:ilp2}), and two Quadratic Unconstrained Binary Optimization (QUBO) problems (See Sections \ref{sec:sQUBO} and \ref{sec:uQUBO}). A QUBO problem is a mathematical program with a quadratic objective function, binary variables, and no explicit constraints. Interestingly, the proposed QUBO problems result from the Constraint Satisfaction Problem (CSP) introduced in Section \ref{sec:CSP}. A CSP is similar to a mathematical program, except that it is not an optimization problem, but rather a satisfaction problem. That is, any set of variables that satisfy the constraints is a valid solution.


Through specialized algorithms based on branch-and-bound, branch-and-cut, branch-and-price, heuristics, and metaheuristics, mathematical programs can be solved or approximated, depending on the difficulty level of each instance. Specialized software known as optimization solvers implement these kind of algorithms allowing us to find optimal or near-optimal solutions. Among the most popular optimization solvers are Gurobi~\cite{gurobi}, CPLEX~\cite{cplex}, SCIP~\cite{scip}, and HiGHS~\cite{highs}. In recent years, quantum-based solvers have emerged as candidates for solving QUBO problems. Therefore, interest in this class of programs has increased. The next section explores this topic.

\subsection{QUBO problems and quantum computing}

QUBO problems are a significant category in mathematical programming because they cover a wide variety of important combinatorial optimization problems~\cite{glover2022quantum}. A QUBO problem is characterized by a quadratic objective function of binary variables with no explicit constraints:
\begin{equation}
    \textbf{min} \ \ \textbf{x}^T\text{Q}\textbf{x} = \sum_i Q_{ii}x_i + \sum_{i<j}Q_{ij}x_ix_j \ \ ,
\label{eq:QUBO}
\end{equation}

where $\textbf{x}\in\{0,1\}^n$ is a vector of $n$ binary decision variables and $\text{Q}$ is an $n\times n$ real-valued matrix. $Q_{ii}$ are the linear coefficients and $Q_{ij}$ are the quadratic coefficients~\cite{kochenberger2014unconstrained}. A constraint in a QUBO model is implicit within the objective function via penalty terms. For example, the explicit constraint $\sum_i y_i = 1$ is equivalent to the implicit constraint $P (\sum_i y_i -1)^2$ if $P$ is sufficiently large. So, by means of penalized implicit constraints, unfeasible solutions are avoided during the search process~\cite{glover2018tutorial}.

As QUBO problems are NP-hard, solving them can be challenging, requiring more specialized approaches. Although optimization solvers such as Gurobi and CPLEX can deal with QUBO problems, they usually struggle to solve large-scale instances~\cite{yang2025novel, kochenberger2014unconstrained}. Thus, for larger or particularly difficult instances, metaheuristics such as Simulated Annealing, Tabu Search, and Genetic Algorithms are employed to find near-optimal solutions~\cite{kochenberger2014unconstrained}. Beyond classical computing, quantum computing has emerged as a valuable alternative. For instance, quantum annealers are physical processors designed to find low-energy states of Ising models, which are mathematically equivalent to QUBO problems~\cite{hua2022adiabatic}. Similarly, gate-based quantum algorithms, such as the Quantum Approximate Optimization Algorithm (QAOA), are used to approximate QUBO solutions on universal quantum computers~\cite{farhi2014quantum}.

In summary, the importance of QUBO models lies in their versatility and broad applicability. From a classical perspective, they provide a unified and often compact framework for representing combinatorial problems. This allows for the application of a wide range of specialized solvers and metaheuristics~\cite{kochenberger2014unconstrained, glover2018tutorial}. Furthermore, the QUBO category is of fundamental importance in quantum computing. It serves as the primary interface for most current quantum optimization approaches in the Noisy Intermediate-Scale Quantum (NISQ) era. For example, by expressing a problem such as the GBP as a QUBO problem, it becomes immediately accessible to quantum annealers and quantum-classical hybrid algorithms such as QAOA, paving the way for new solving techniques~\cite{de2024optimized, glover2022quantum, glover2022quantum2}.

\section{Related work}
\label{sec2}
As there is no exact formula for determining the burning number of arbitrary graphs, a significant amount of research has been directed towards finding tighter upper bounds for the burning number in both arbitrary and particular graph families. Bonato et al.~\cite{bonato2016burn} proposed the Burning Number Conjecture, which states that for any connected graph $G$ of order $n$, $b(G)\leq \lceil \sqrt{n} \rceil$. In fact, $b(P_n)=b(C_n)= \lceil \sqrt{n} \rceil$, where $P_n$ and $C_n$ are paths and cycles of order $n$, respectively~\cite{bonato2014burning}. This conjecture also holds for spiders~\cite{bonato2019bounds}, caterpillars~\cite{liu2020burning, hiller2020burning}, homeomorphically irreducible trees~\cite{murakami2023graph}, and graphs with minimum degree $\delta \geq 23$~\cite{kamali2020burning}. Trivial cases include complete graphs $K_n$ and star graphs $K_{1,n}$ where $b(K_n)=b(K_{1,n})=2$. Regarding general upper bounds, the best known bound is due to Bonato and Kamali, who proved $b(G)\leq \big \lceil \frac{\sqrt{12n+64}+8}{3} \big\rceil$~\cite{bonato2021improved}, improving earlier results by Land and Lu~\cite{land2016upper}. Further progress has been made for other graph families, such as generalized Petersen graphs~\cite{sim2018burning}, theta graphs~\cite{liu2019burning}, random graphs~\cite{mitsche2017burning}, grids~\cite{mitsche2017burning, mitsche2018burning}, hypercubes~\cite{li2024characterization}, trees~\cite{das2023burning}, and path forests~\cite{bonato2019bounds, liu2021burning}.

Beyond the classical GBP, several extensions and variations have been proposed throughout the years to better model and study real-world propagation dynamics. The first extension of GBP was proposed by Mitsche et al.~\cite{mitsche2017burning} aimed at exploring the probabilistic aspects of the problem. Later, it was extended to specific graph settings such as geometric graphs~\cite{gupta2020burning}, hypergraphs~\cite{burgess2024extending}, temporal graphs~\cite{enright2024structural} and directed graphs~\cite{janssen2020burning}, and to metric spaces including anywhere burning~\cite{keil2022burning} and euclidean burning~\cite{chandarana2024graph}. Regarding the variants of GBP, the first two proposed were $k$-fast, where the fire can spread at a distance of at most $k$~\cite{moghbel2020topics}, and $k$-slow burning, where the fire spreads to at most $k$ neighbors~\cite{moghbel2020topics}. Subsequent work includes edge burning~\cite{mondal2021apx, antony2024graph}, where the edges are burned instead of the vertices, total burning, where both edges and vertices are burned~\cite{moghbel2020topics, antony2024graph}, generalized burning number, where a vertex can only be burned if at least $r$ neighbors were burned in the previous round~\cite{li2021generalized}, $w$-burning, where $w$ vertices can be ignited spontaneously each round~\cite{mondal2021apx}, and $\theta$-GBP, where a threshold $\theta$ is assigned to each vertex that limits how many neighbors each can burn~\cite{Iurlano2024ConstrainedDiffusion}. Other variants include game-theoretic formulations such as Burning Game~\cite{chiarelli2024burning} and Adversarial Burning~\cite{gunderson2025adversarial}, stochastic versions such as the Independent Cascade Model of Graph Burning~\cite{song2023independent}, and color-constrained models such as Chromatic Burning and Edge Chromatic Burning~\cite{Komala2024BurningEdgeChromatic}.

Given its NP-hard nature, it seems unlikely that the GBP can be solved in polynomial time. Nevertheless, many efficient heuristics, metaheuristics, and approximation algorithms have been designed. Among the most relevant are a 2-approximation algorithm for trees~\cite{bonato2019approximation}, a 1.5-approximation algorithm for graphs with disjoint paths~\cite{bonato2019approximation}, a 2-approximation algorithm for square grids~\cite{gupta2021burning}, and $(3-2/b(G))$-approximation algorithms for arbitrary graphs~\cite{bonato2019approximation,garcia2022burning}. While most heuristics and metaheuristics are based on centrality measures such as eigenvector and betweenness centrality~\cite{nazeri2023centrality,simon2019burn,gautam2022faster,vsimon2019heuristics}, a more recent one is based on a 0.5-approximation algorithm for the CMCP~\cite{garcia2025greedy}. Although these heuristics and metaheuristics are good at finding near-optimal solutions, they do not give any optimality guarantee. With regard to mathematical programs, many have been previously reported~\cite{garcia2025greedy,garcia2022graph,pereira_et_al:LIPIcs.ESA.2024.94}. Inspired by them, this paper reports mathematical programs that go from subtle enhancements of those previously known to completely novel ones.


\section{Mathematical programs for the graph burning problem}
\label{sec3}

This section introduces novel mathematical programs for the GBP. These are one MILP (see Section \ref{sec:milp}), one CSP (see Section \ref{sec:CSP}), two ILPs (see Sections \ref{sec:ilp1} and \ref{sec:ilp2}), and two QUBO problems (see Section \ref{sec:sQUBO} and \ref{sec:uQUBO}).

\subsection{A MILP based on the propagation process: PROP-MILP}
\label{sec:milp}

\begin{figure}[h]
    \centering
    \includegraphics[width=0.75\linewidth]{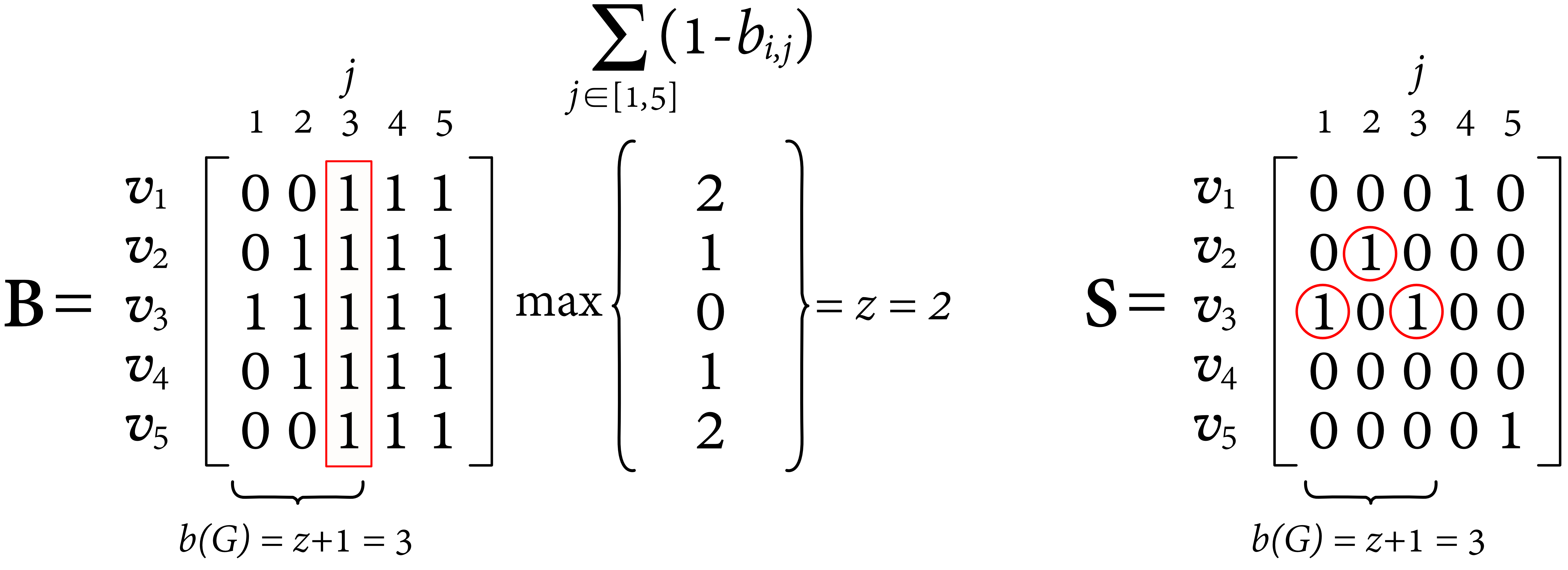}
    \caption{An optimal solution, $(v_3,v_2,v_3)$, to PROP-MILP. The input graph is $P_5$ (see Figure \ref{fig:2}), $\textbf{S}=[s_{i,j}]_{n \times U}$, $\textbf{B}=[b_{i,j}]_{n \times U}$, and $U=5$. Notice how $\textbf{B}$ correctly codifies the propagation process}
    \label{fig:4}
\end{figure}

Expressions (\ref{MP:1:1}) to (\ref{MP:1:6}) define PROP-MILP, a MILP for the GBP with a known upper bound $U$ for $b(G)$. This program is inspired by the ILP for the Firefighter Problem (FP) \cite{finbow2009firefighter} and is based on an ILP for the GBP reported in \cite{garcia2025greedy}. Assuming that $V(G)=\{v_1,v_2,...,v_n\}$, the binary variables $s_{i,j}$ and $b_{i,j}$ codify the burning sequence and the burning process, respectively. Vertices cannot be burned in more than $z$ steps (see Expression (\ref{MP:1:4})), a value that is minimized (see Expression (\ref{MP:1:1})). Although $z$ is an integer, for practical purposes it is better to set it as a real value. Vertices can catch fire only if they are in the sequence or have a previously burned neighbor (see Expression (\ref{MP:1:2})). Finally, only one vertex can be added to the sequence at each round of the propagation process (See Expression (\ref{MP:1:3})). Notice that the solution is codified into variables $s_{i,j}$ and its length is $b(G)=z+1$ (see Figure \ref{fig:4}).

\begin{align}
	\textbf{min} & \ \ z &  \label{MP:1:1} \\
	\textbf{s.t.}
	& \ \ b_{i,j} \le s_{i,j} + \sum \limits_{v_k\in N[v_i]} b_{k,j-1} &  \forall v_i \in V, \forall j \in [1,U]  \label{MP:1:2}\\
	& \ \ \sum \limits_{v_i \in V} s_{i,j} = 1 &   \forall j \in [1,U]  \label{MP:1:3}\\
	& \ \ \sum \limits_{j\in[1,U]} (1-b_{i,j})  \le z &   \forall v_i \in V  \label{MP:1:4}\\  
	\textbf{where}
	& \ \ s_{i,j} \ , \ b_{i,j} \in \{0,1\} \ , \ b_{i,0}=0  &   \forall v_i \in	V, \forall j \in [1,U]  \label{MP:1:5}\\
    & \ \ z \in \mathbb{R} \ , \ b(G)\le U \label{MP:1:6}
\end{align}

\begin{figure}[h]
    \centering
    \includegraphics[width=0.5\linewidth]{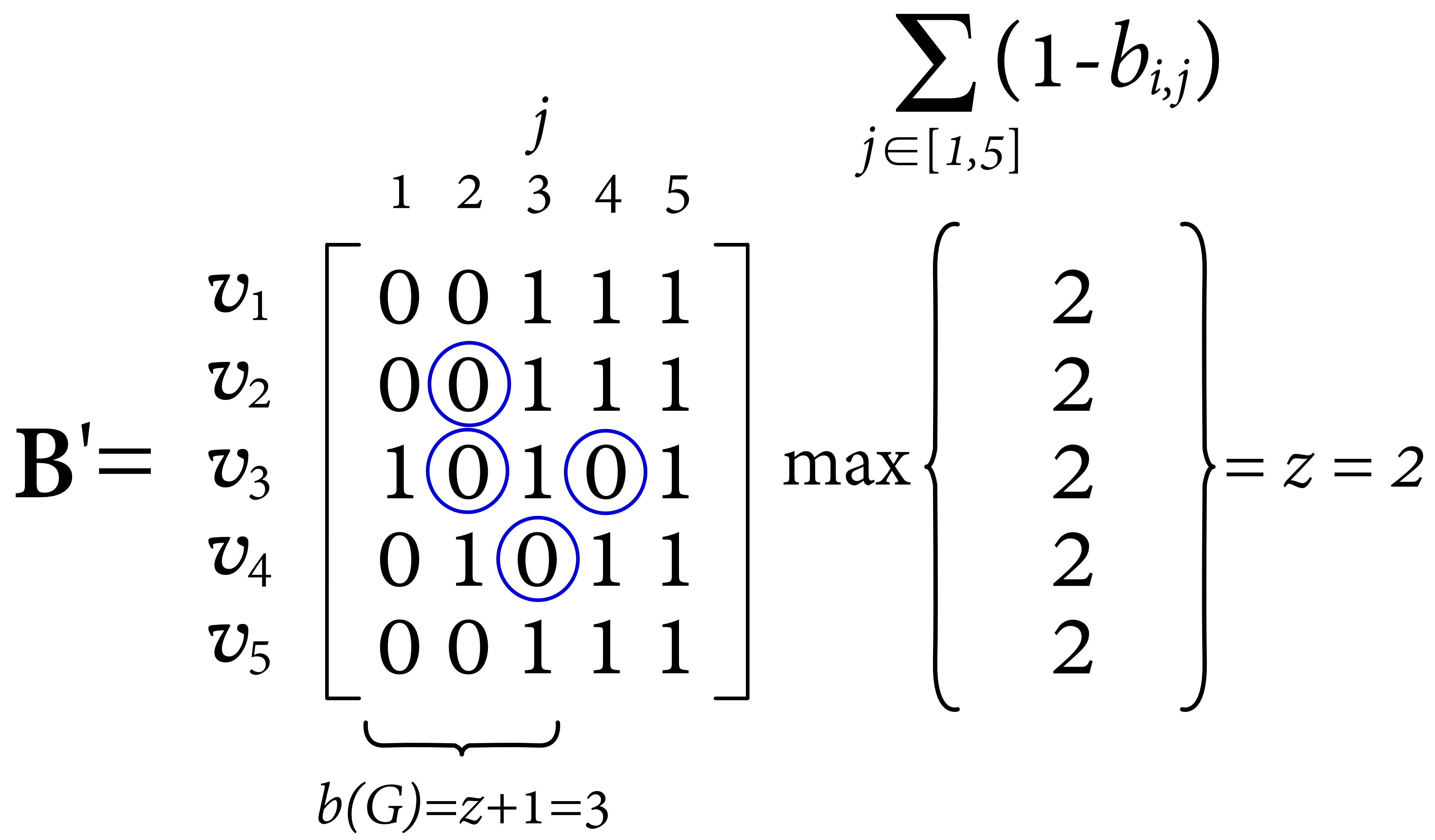}
    \caption{Although matrix $\textbf{B}'$ does not correctly codify the burning process, it is as valid as $\textbf{B}$ from the previous figure}
    \label{fig:5}
\end{figure}

From all the programs presented in this paper, PROP-MILP has more variables, $2Un$. However, it is perhaps the most intuitive. Hence, we decided to present it first. As its name suggests, it is based on the propagation process. As Figures \ref{fig:4} and \ref{fig:5} show, the propagation process may not be accurately represented by the variables $b_{i,j}$. However, the optimal burning sequence is correctly codified in variables $s_{i,j}$. In detail, for the input graph $P_5$ and $U=5$, the optimal solution to PROP-MILP codifies the optimal burning sequence in matrix $\textbf{S}=[s_{i,j}]_{n \times U}$. However, matrices $\textbf{B}=[b_{i,j}]_{n \times U}$ and $\textbf{B}'=[b'_{i,j}]_{n \times U}$ are equally valid even if only $\textbf{B}$ correctly codifies the propagation process, i.e., the circled zeros in $\textbf{B}'$ should be ones.

The main drawback of PROP-MILP is its large number of variables. The more variables, the larger the search space, which can significantly increase the execution time of exact algorithms. Fortunately, all subsequent formulations use fewer variables than PROP-MILP, making them more practical alternatives.

\subsection{A CSP based on the CMCP: COV-CSP}
\label{sec:CSP}

Expressions (\ref{MP:2:1}) to (\ref{MP:2:4}) define COV-CSP, a CSP for the GBP with a guess $g$ on $b(G)$. This program is based on an ILP reported in \cite{garcia2025greedy}. As its name suggests, COV-CSP leverages the reduction from the GBP to the CMCP.

\begin{figure}[h]
    \centering
    \includegraphics[width=0.47\linewidth]{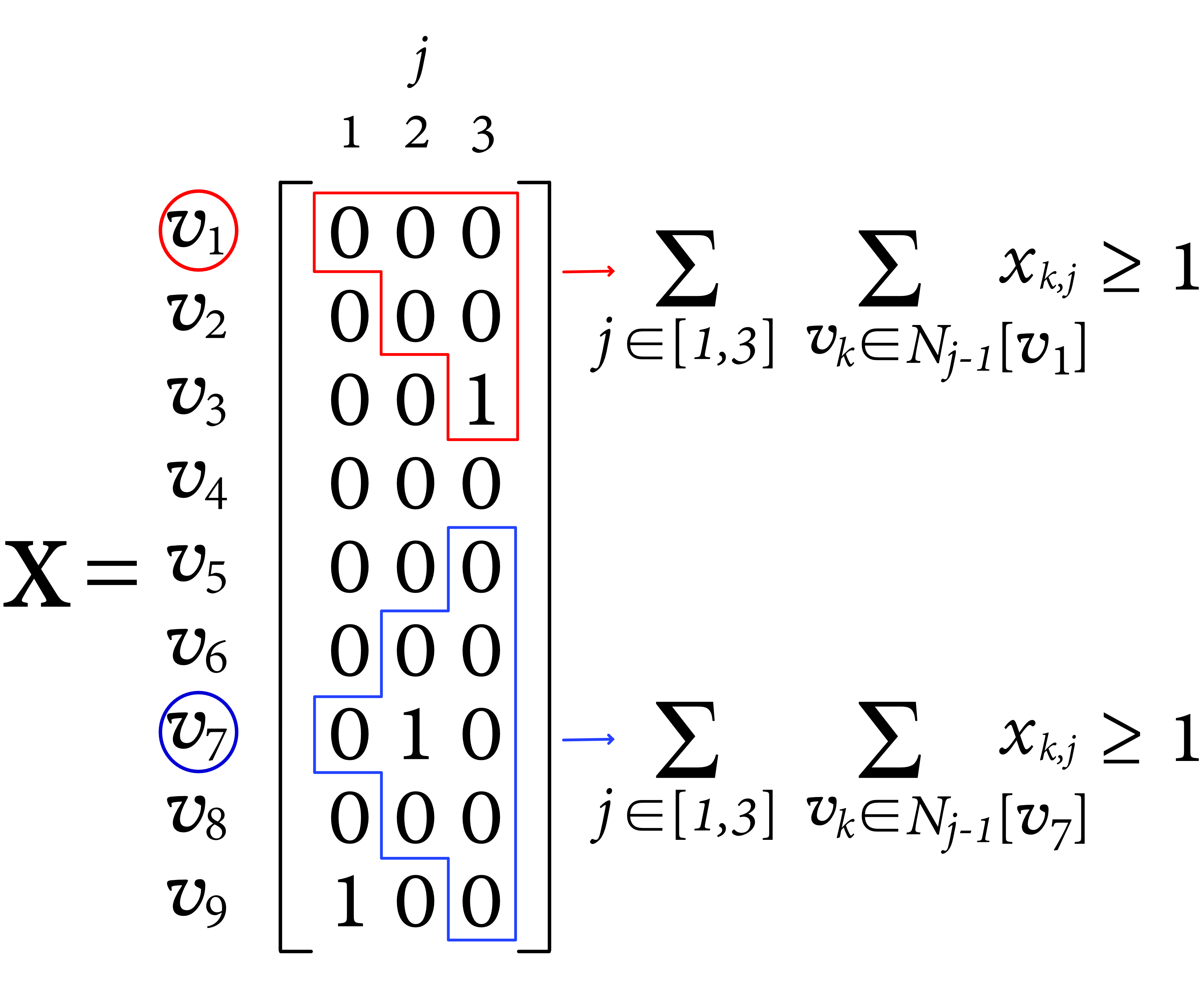}
    \caption{An optimal solution, $(v_3,v_7,v_9)$, to COV-CSP with input graph $P_9$ and $g=b(P_9)=3$. Note that all double sums have a value of at least one}
    \label{fig:6}
\end{figure}

First, $j$ represents the available \textit{coverage radii}, which go from $0$ to $g-1$. In more detail, Expression (\ref{MP:2:2}) guarantees that only one vertex for each \textit{coverage radius} is selected, and Expression (\ref{MP:2:3}) ensures that all vertices are \textit{burned} by at least one vertex in the burning sequence. Figure \ref{fig:6} shows an optimal solution of COV-CSP with input graph $P_9$ and $g=b(P_9)=3$.

\begin{align}
	\textbf{find} & \ \ [x_{i,j}]_{n\times g} &  \label{MP:2:1} \\
	\textbf{s.t.} & \ \ \sum\limits_{v_i \in V} x_{i,j}=1 &   \forall j \in [1,g]  \label{MP:2:2}\\
	& \ \ 1\le \sum\limits_{j\in [1,g]} \ \sum_{v_k \in N_{j-1}[v_i]} x_{k,j} &   \forall v_i \in V \label{MP:2:3}\\
	\textbf{where}
	& \ \ x_{i,j} \in \{0,1\}  &  \forall v_i \in V, \forall j \in [1,g]  \label{MP:2:4}
\end{align}

The main advantage of COV-CSP is its reduced number of variables, $gn$. However, it has the disadvantage of requiring $b(G)$ to be known in advance. To address this issue, it can be embedded in a binary search. That is, lines 5 and 6 from Algorithm \ref{alg:1} can be replaced by 

\begin{equation*}
S \Leftarrow \text{Solve COV-CSP} (G,g)  
\end{equation*}

and 

\begin{equation*}
\textbf{if } \text{the problem is feasible} \textbf{ then} \ \ ,
\end{equation*}

repectively. Since $g < b(G)$ makes COV-CSP infeasible, this approach has the disadvantage of requiring an infeasibility test, which could be very time-consuming. To avoid this issue, COV-CSP can be replaced by the ILP presented in Section \ref{sec:ilp1}. 


\subsection{An ILP based on the CMCP: COV-ILP}
\label{sec:ilp1}

\begin{figure}[h]
    \centering
    \includegraphics[width=0.8\linewidth]{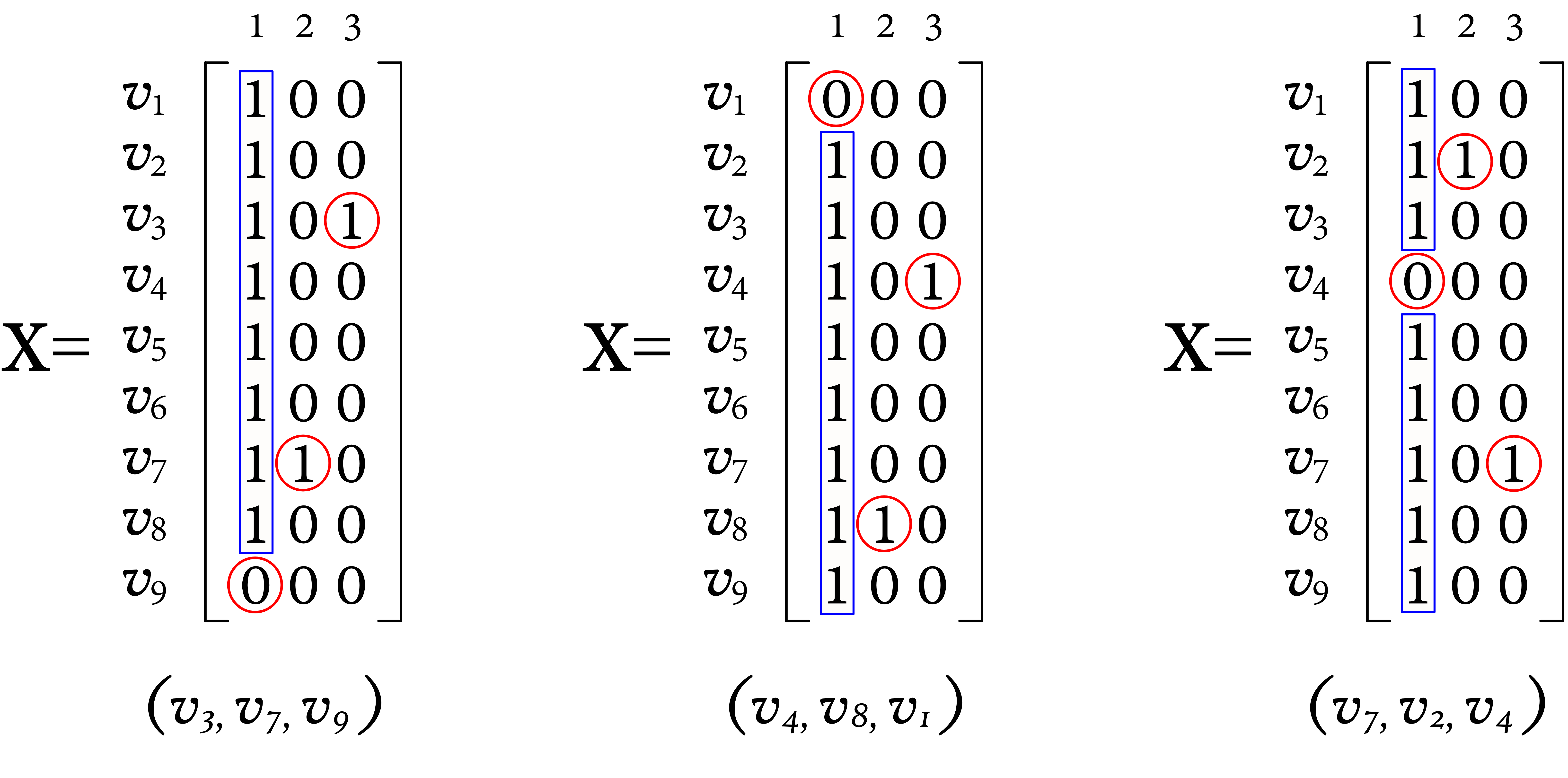}
    \caption{Three different optimal solutions for the COV-ILP with input graph $P_9$ and $g= b(P_9)=3$}
    \label{fig:7}
\end{figure}

Expressions (\ref{MP:3:1}) to (\ref{MP:3:4}) define COV-ILP, an ILP for the GBP. Like COV-CSP, this program leverages the relationship between the GBP and the CMCP, and improves an ILP reported in \cite{garcia2025greedy}. Expression (\ref{MP:3:2}) guarantees that only one vertex for each \textit{coverage radius} is selected, except for $0$, which corresponds to $j=1$. Expression (\ref{MP:3:3}) along with the objective function (\ref{MP:3:1}) seeks to \textit{burn} as many vertices as possible. Note that the variables $x_{i,1}$ are used to count the number of \textit{burned} vertices. Thus, if 

\begin{equation}
\sum_{v_i \in V} x_{i,1}=n \ \ ,    
\end{equation}

then any vertex can be selected as the last vertex in the burning sequence. If 

\begin{equation}
\sum_{v_i \in V} x_{i,1}=n-1 \ \ ,
\end{equation}

then the vertex $v_i$, such that $x_{i,1}=0$, must be the last vertex in the burning sequence (see Figure \ref{fig:7}).

\begin{align}
	\textbf{max} & \ \ \sum\limits_{v_i \in V} x_{i,1}  &  \label{MP:3:1} \\
	\textbf{s.t.} & \ \ \sum\limits_{v_i \in V} x_{i,j}=1 &   \forall j \in [2,g]  \label{MP:3:2}\\
	& \ \ x_{i,1}\le \sum\limits_{j\in [2,g]} \ \sum_{v_k \in N_{j-1}[v_i]} x_{k,j} &   \forall v_i \in V \label{MP:3:3}\\
	\textbf{where}
	& \ \ x_{i,j} \in \{0,1\} &  \forall v_i \in V, \forall j \in [1,g]  \label{MP:3:4}
\end{align}

While $g < b(G)$ makes COV-CSP infeasible, COV-ILP is always feasible regardless of the value of $g$. Thus, COV-ILP seems to be better suited for solving the GBP embedded in a binary search. To do so, lines 5 and 6 of Algorithm \ref{alg:1} must be replaced by 

\begin{equation*}
S \Leftarrow \text{Solve COV-ILP} (G,g) 
\end{equation*}

and

\begin{equation*}
\textbf{if } S \text{ is a burning sequence} \textbf{ then} \ \ ,
\end{equation*}

respectively. So far, we have defined three mathematical programs for the GBP: PROP-MILP, COV-CSP, and COV-ILP. Although PROP-MILP is perhaps the most intuitive, it is also the one with the most variables, $2Un$. COV-CSP is simpler, has only $gn$ variables, but requires knowing $b(G)$ in advance. Besides, if $g<b(G)$, COV-CSP is infeasible. COV-ILP improves COV-CSP in the sense that it is always feasible regardless of the value of $g$. To be practical, both COV-CSP and COV-ILP must be embedded in a binary search; thus, they must be solved up to $\log_2 n$ times. The following ILP, which has $Un$ variables, is better than the previous programs in the sense that it does not require being embedded in a binary search.

\subsection{An ILP based on the CMCP with known upper bound: GBP-ILP}
\label{sec:ilp2}

\begin{figure}[h]
    \centering
    \includegraphics[width=0.7\linewidth]{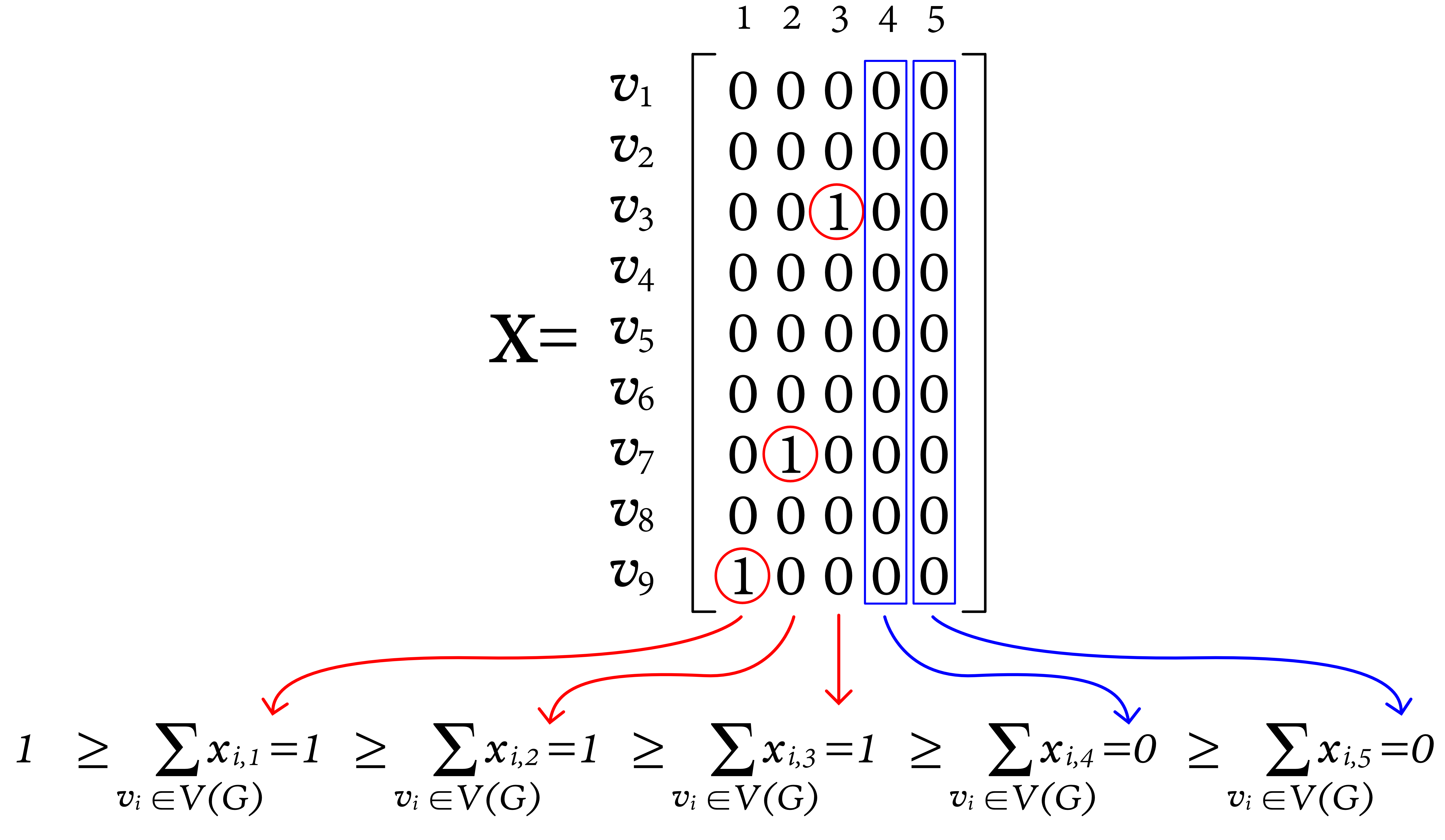}
    \caption{An optimal solution for GBP-ILP with input graph $P_9$ and $b(P_9)=3\le U=5$}
    \label{fig:8}
\end{figure}

Expressions (\ref{MP:4:1}) to (\ref{MP:4:5}) define GBP-ILP, an ILP for the GBP. This program is an enhancement of an ILP reported in~\cite{garcia2025greedy}. Unlike COV-CSP and COV-ILP, which require a guess $g$ on $b(G)$, GBP-ILP only requires an upper bound $U$ on $b(G)$. Expression (\ref{MP:4:2}) guarantees that only one vertex for each \textit{coverage radius} is selected and that these are increasing non-negative numbers starting at 0. Expression (\ref{MP:4:3}) guarantees that all vertices are burned. Figure \ref{fig:8} shows an optimal solution for GBP-ILP with input graph $P_9$.

\begin{align}
	\textbf{min} & \ \ \sum\limits_{j \in [1,U]} \sum\limits_{v_i \in V} x_{i,j} &  \label{MP:4:1} \\
	\textbf{s.t.} & \ \ \sum_{v_i \in V}x_{i,j} \le \sum_{v_i \in V}x_{i,j-1} \le 1 & \forall j \in [2,U] \label{MP:4:2}\\
	& \ \ 1 \le \sum\limits_{j\in [1,U]} \ \sum_{v_k \in N_{j-1}[v_i]} x_{k,j} &   \forall v_i \in V \label{MP:4:3}\\
	\textbf{where}
	& \ \ x_{i,j} \in \{0,1\} &  \forall v_i \in V, \forall j \in [1,U]   \label{MP:4:4}\\
        & \ \ b(G)\le U \label{MP:4:5}
\end{align}

So far, it might seem like PROP-MILP, COV-CSP, and COV-ILP are somehow inferior to GBP-ILP. To some extent, this is true because we found Gurobi to be capable of solving GBP-ILP on instances with millions of vertices using a row generation method similar to the one used by Pereira et al.~\cite{pereira_et_al:LIPIcs.ESA.2024.94} (see Appendix~\ref{secA1}). However, this does not undermine the potential usefulness of the other programs. For instance, to define the following QUBO problems (see Sections \ref{sec:sQUBO} and \ref{sec:uQUBO}), we used COV-CSP as a basis.

\subsection{A QUBO problem with slack variables: sQUBO}
\label{sec:sQUBO}

In this section, we define sQUBO, which results from embedding the constraints of COV-CSP into the objective function. For the equality constraints (\ref{MP:2:2}), we use the following observation:

\begin{equation}
    y_1+y_2=1 \ \ \ \ \iff \ \ \ \ 1-y_1-y_2+2y_1y_2=0 \ ,
\end{equation}

which generalizes to any number of variables,

\begin{equation}
     \sum_{1=1}^{m}y_i=1 \ \ \ \ \iff \ \ \ \ 1-\sum_{i=1}^{m}y_i+2\sum_{i=1}^{m-1}\sum_{j=i+1}^{m}y_iy_j=0 \ .
    \label{eq:equality}
\end{equation}

In more detail, the right part of Expression (\ref{eq:equality}) evaluates to $1$ when all variables are set to 0, evaluates to $0$ when only one variable is set to 1, and evaluates to a positive number when two or more variables are set to 1. Thus, the minimum value of 0 is achieved by setting exactly one variable to 1.\\

To incorporate Constraints (\ref{MP:2:3}) into the objective function, we used the following observation.

\begin{equation}
    \begin{aligned}
        \sum_{i=1}^m y_i \ge 1  \ \ \ \ \iff \ \ \ \ & \left( 1- \sum_{i=1}^{m}y_i + \sum_{i=1}^{m'}2^{i-1}s_i \right)^2 = 0 \ .
    \end{aligned}
    \label{eq:inequality}
\end{equation}

By adding slack variables $s_i$, the right part of Expression (\ref{eq:inequality}) has a value of 0 when the left part is satisfied. Since the right part is squared, 0 is its minimum value. Using observations (\ref{eq:equality}) and (\ref{eq:inequality}), we get the following QUBO problem for GBP, which we refer to as sQUBO.

\begin{align}
    \textbf{min} & \ \ \sum\limits_{j\in[1,g]} \left( 1-\sum\limits_{v_i \in V}x_{i,j} + 2\sum_{v_i \in V} \ \sum_{k \in [i+1,n]} x_{i,j}x_{k,j} \right)&  \label{MP:5:1}\\
    & \ \  + \sum\limits_{v_i \in V} \Bigg( 1 -\sum\limits_{j\in[1,g]} \ \sum\limits_{v_k \in N_{j-1}[v_i]} x_{k,j} + \sum_{\ell\in [1,\lceil \log_2 g \rceil]} 2^{\ell-1} s_{i,\ell} \Bigg)^2 \  &  \label{MP:5:2}\\
    \textbf{where}
	& \ \ x_{i,j} \in \{0,1\} \ \ \ \ \ \forall v_i \in V, \forall j \in [1,g]  \label{MP:5:3}
\end{align}

The number of slack variables per vertex, $\lceil \log_2 g \rceil$, is because only one vertex must be selected for each available coverage radius. Notice that if $g\ge b(G)$, all vertices must be covered; thus, the value of the optimal solution is $0$. In case $g<b(G)$, the optimal solution size must be greater than 0. That is, while $g<b(G)$ leads to an infeasible COV-CSP problem, sQUBO is feasible regardless of the value of $g$. In this way, sQUBO can be embedded in a binary search. That is, lines 5 and 6 from Algorithm \ref{alg:1} should be replaced by 

\begin{equation*}
S \Leftarrow \text{Solve sQUBO}(G,g)
\end{equation*}

and

\begin{equation*}
\textbf{if } OPT=0 \textbf{ then} \ \ ,
\end{equation*}

respectively. Although sQUBO correctly models the GBP, it has the disadvantage of requiring additional slack variables, which increases the size of the search space. To address this issue, we used an unbalanced penalization technique in the following section.

\subsection{A QUBO problem with unbalanced penalization: uQUBO}
\label{sec:uQUBO}

\begin{figure}[h]
    \centering
    \includegraphics[width=0.8\linewidth]{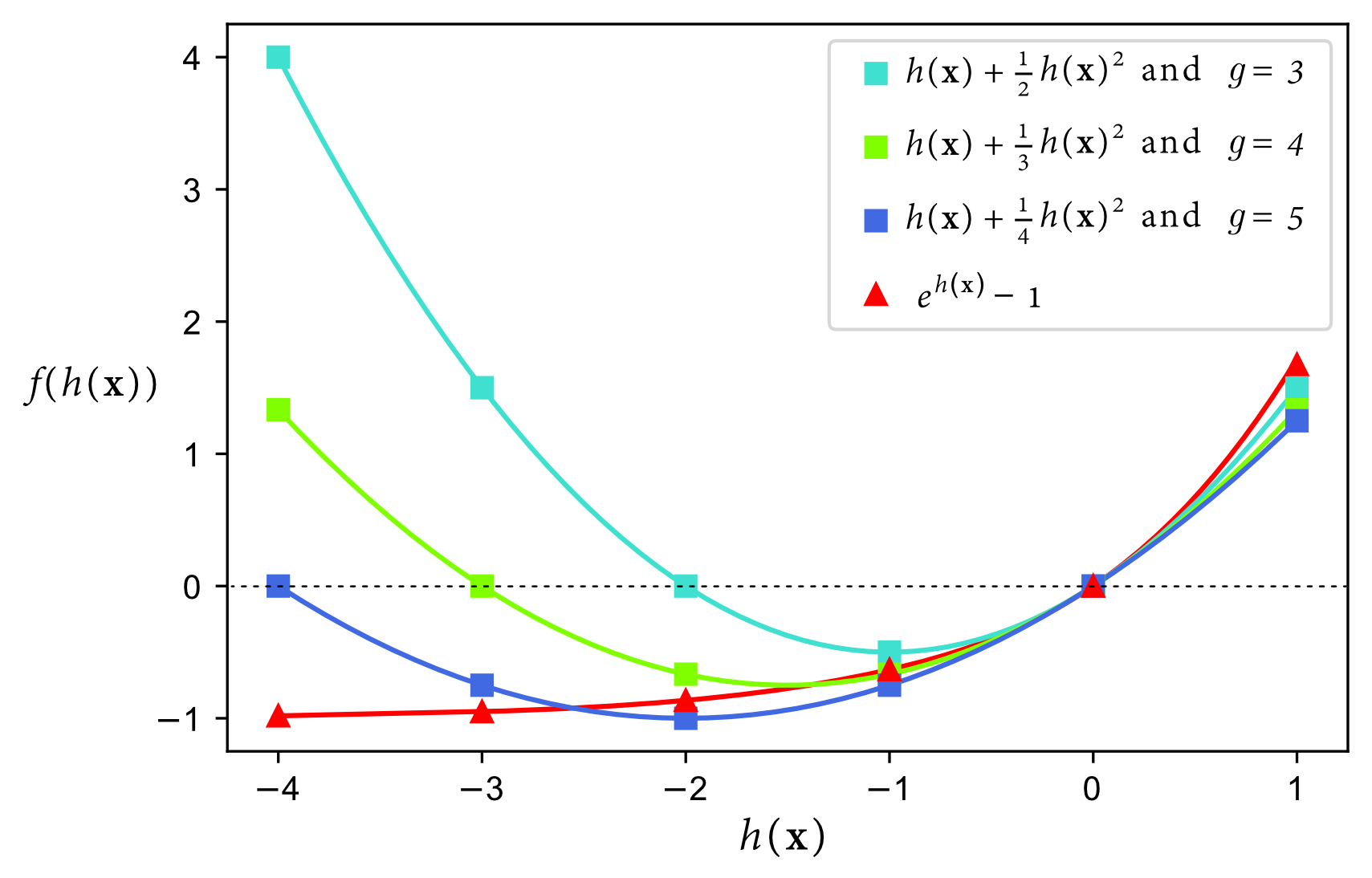}
    \caption{Different approximations to $e^{h(\textbf{x})}-1$}
    \label{fig:9}
\end{figure}

The unbalanced penalization technique reported in \cite{montanez2024unbalanced} models inequality constraints of the form

\begin{equation}
    h(\textbf{x}) = c - \sum_{i} \ell_i x_i \le 0
\end{equation}

with the approximation

\begin{equation}
    e^{h(\textbf{x})} - 1 \approx \lambda_a h(\textbf{x}) + \lambda_b h(\textbf{x})^2 = f(h(\textbf{x})) \ \ ,
\end{equation}

where $\lambda_a$ and $\lambda_b$ are penalization terms that can be fine-tuned for each specific problem (see Figure \ref{fig:9}). Constraints (\ref{MP:2:3}) from COV-CSP can be incorporated into the objective function as Expressions (\ref{MP:6:2}) and (\ref{MP:6:3}). We refer to the resulting QUBO problem as uQUBO. Thanks to unbalanced penalization, slack variables are no longer required. However, as we show later, there is a price to pay.

\begin{align}
    \textbf{min} & \ \ P \sum\limits_{j\in [1,g]} \left( 1-\sum\limits_{v_i \in V}x_{i,j} + 2\sum_{v_i \in V} \ \sum_{k\in[i+1,n]} x_{i,j}x_{k,j} \right)&  \label{MP:6:1}\\
    & \ \  + \lambda_1 \sum\limits_{v_i\in V} \Bigg( 1 -\sum\limits_{j\in[1,g]} \ \sum\limits_{v_k \in N_{j-1}[v_i]} x_{k,j} \Bigg)  \  &  \label{MP:6:2}\\
    & \ \  + \sum\limits_{v_i\in V} \lambda_{2,i} \Bigg( 1 -\sum\limits_{j\in[1,g]} \  \sum\limits_{v_k \in N_{j-1}[v_i]} x_{k,j} \Bigg)^2   &  \label{MP:6:3}\\
        \textbf{where}
	& \ \ x_{i,j} \in \{0,1\} \ \ \ \ \ \forall v_i \in V, \forall j \in [1,g]  \label{MP:6:4}
\end{align}

The penalization terms $P$, $\lambda_1$, and each $\lambda_{2,i}$ must be properly tuned. First, we must ensure that having $1,2,$ and up to $g$ \textit{fire sources} for each vertex is less penalized than any other option, where $g$ is our guess on $b(G)$. Thus, we must set $\lambda_{2,i} = \frac{\lambda_1}{g-1}$ (see Eq. (\ref{eq:lambdas})). In other words, setting $\lambda_{2,i} = \frac{\lambda_1}{g-1}$ guarantees that a vertex having $1,2,$ and up to $g$ \textit{fire sources} contributes to the value of the objective function with a non-positive value $f(h(\textbf{x}))$ (see Figure \ref{fig:9}).

\begin{equation}
    f(1-g) = \lambda_1 (1-g) + \lambda_{2,i} (1-g)^2 = 0 \ \ .
    \label{eq:lambdas}
\end{equation}

Second, we must guarantee that Expression (\ref{MP:6:1}) has a sufficiently large penalization $P$; otherwise, the solution might not be a valid burning sequence. The value of $P$ must be higher than the absolute value of the smallest possible value of Expressions (\ref{MP:6:2}) and (\ref{MP:6:3}). In this way, the size of the optimal solution to uQUBO must be at most zero, which implies that all constraints modeled by expression (\ref{MP:6:1}) are satisfied. Since

\begin{equation}
    \frac{d(\lambda_1h(\textbf{x})+\lambda_2 h(\textbf{x})^2)}{dh(\textbf{x})} = \lambda_1+2\lambda_2 h(\textbf{x}) \ \ ,
\end{equation}

the smallest possible value Expressions (\ref{MP:6:2}) and (\ref{MP:6:3}) can take for each vertex is $h(\textbf{x})=\frac{-\lambda_1}{2\lambda_2}$. Namely, 

\begin{equation}
    f_{min} = f\Bigl(\frac{-\lambda_1}{2\lambda_2}\Bigr) = -\frac{\lambda_1^2}{4\lambda_2} \ \ .
\end{equation}

Thus, $P > |nf_{min}|$ ensures Expression (\ref{MP:6:1}) to evaluate to zero. Third, Expressions (\ref{MP:6:2}) and (\ref{MP:6:3}) lead to a situation where many vertices with many \textit{fire sources} might be better than having all vertices \textit{burned}. If the number of \textit{fire sources} per vertex is known in advance, we could fine-tune each $\lambda_{2,i}$, that is, we could set the exact number of \textit{fire sources} each vertex must have. Unfortunately, this cannot be done since it requires knowing an optimal solution in advance. However, we can use a feasible solution as a guide. In this manner,

\begin{equation}
\lambda_{2,i} =
     \begin{cases} 
      \frac{\lambda_1}{\ell_i - 1} & \text{if} \ \ \ell_i \ge 2 \\
      \ \ \lambda_1 & \text{if} \ \ \ell_i = 1
   \end{cases} \ \ ,
   \label{eq:feasible}
\end{equation}

where $\ell_i$ is the number of \textit{fire sources} of each $v_i \in V(G)$ for a given burning sequence. A heuristic or approximated algorithm can be used to efficiently construct such a sequence. Finally, to embed uQUBO into a binary search, lines 5 and 6 from Algorithm \ref{alg:1} should be replaced by 

\begin{equation*}
S \Leftarrow \text{Solve uQUBO}(G,g) 
\end{equation*}

and 

\begin{equation*}
\textbf{if } S \text{ is a burning sequence } \textbf{then} \ \ ,
\end{equation*}

respectively. Nevertheless, the optimal solution to uQUBO is not guaranteed to be a valid burning sequence; this is the cost of not having slack variables. To assess how often uQUBO yields optimal burning sequences, we solved it using Gurobi 12.0.3~\cite{gurobi} and tested it on Erdős–Rényi and geometric random graphs. The results are reported in Tables~\ref{tab:erdos} and \ref{tab:geo}. In both cases, we compared $\lambda_{2,i}=1/(g-1)$ and $\lambda_{2,i}=1/(\ell_i-1)$. For all experiments, we set $\lambda_1=1$, and the guiding feasible solution was provided by the polynomial-time greedy algorithm, Gr, from~\cite{garcia2025greedy}. As these tables show, $\lambda_{2,i}=1/(g-1)$ led to poor performance, particularly in graphs with many connected components, whereas $\lambda_{2,i}=1/(\ell_i-1)$ enabled uQUBO to successfully solve all generated instances. Since QUBO problems are difficult to solve in classical optimization solvers, we solved only small instances.


\begin{table}[h]
\caption{Percentage of optimal burning sequences found by Gurobi, using uQUBO embedded into a binary search. One hundred Erd\H{o}s-Renyi graphs were generated for each pair of parameters $(n,p)$; $\langle c\rangle$ represents the average number of connected components.}\label{tab2}
\begin{tabular*}{\textwidth}{@{\extracolsep\fill}crccccccccc}
\toprule%
\multirow{2}{*}{$p$}&$n$ & \multicolumn{3}{@{}c@{}}{$9$} & \multicolumn{3}{@{}c@{}}{$12$} & \multicolumn{3}{@{}c@{}}{$15$} \\
\cmidrule{3-5}\cmidrule{6-8}\cmidrule{9-11}
& $\lambda_{2,i}$ & $\frac{1}{g-1}$ & $\frac{1}{\ell_i-1}$ & $\langle c \rangle$ & $\frac{1}{g-1}$ & $\frac{1}{\ell_i-1}$ & $\langle c \rangle$ & $\frac{1}{g-1}$ & $\frac{1}{\ell_i-1}$ & $\langle c \rangle$\\
\midrule
$1/2n$ && 13\%  & \textbf{100\%} & 7.05 & 6\%  & \textbf{100\%} & 9.10 & 2\%  & \textbf{100\%} & 11.37\\
$1/n$  && 14\%  & \textbf{100\%} & 5.05 & 2\%  & \textbf{100\%} & 6.73 & 2\%  & \textbf{100\%} & 7.92\\
$3/2n$ && 48\%  & \textbf{100\%} & 3.49 & 26\% & \textbf{100\%} & 4.39 & 15\% & \textbf{100\%} & 5.13\\
$2/n$  && 73\%  & \textbf{100\%} & 2.38 & 58\% & \textbf{100\%} & 2.73 & 42\% & \textbf{100\%} & 3.25\\
$5/2n$ && 84\%  & \textbf{100\%} & 1.73 & 79\% & \textbf{100\%} & 1.88 & 70\% & \textbf{100\%} & 2.20\\
$3/n$  && 95\%  & \textbf{100\%} & 1.29 & 90\% & \textbf{100\%} & 1.45 & 88\% & \textbf{100\%} & 1.63\\
$7/2n$ && 98\%  & \textbf{100\%} & 1.13 & 94\% & \textbf{100\%} & 1.25 & 90\% & \textbf{100\%} & 1.28\\
$4/n$  && \textbf{100\%} & \textbf{100\%} & 1.06 & 99\% & \textbf{100\%} & 1.11 & 98\% & \textbf{100\%} & 1.16\\
$9/2n$  && \textbf{100\%} & \textbf{100\%} & 1.02 & \textbf{100\%} & \textbf{100\%} & 1.07 & 98\% & \textbf{100\%} & 1.10\\
$5/n$  && \textbf{100\%} & \textbf{100\%} & 1.01 & \textbf{100\%} & \textbf{100\%} & 1.03 & \textbf{100\%} & \textbf{100\%} & 1.04\\
\midrule
\end{tabular*}
\label{tab:erdos}
\end{table}

\begin{table}[h]
\caption{Percentage of optimal burning sequences found by Gurobi, using uQUBO embedded into a binary search. One hundred geometric random graphs were generated for each pair of parameters $(n,r)$; $\langle c\rangle$ represents the average number of connected components.}\label{tab2}
\begin{tabular*}{\textwidth}{@{\extracolsep\fill}crccccccccc}
\toprule%
\multirow{2}{*}{$r$}&$n$ & \multicolumn{3}{@{}c@{}}{$9$} & \multicolumn{3}{@{}c@{}}{$12$} & \multicolumn{3}{@{}c@{}}{$15$} \\
\cmidrule{3-5}\cmidrule{6-8}\cmidrule{9-11}
& $\lambda_{2,i}$ & $\frac{1}{g-1}$ & $\frac{1}{\ell_i-1}$ & $\langle c \rangle$ & $\frac{1}{g-1}$ & $\frac{1}{\ell_i-1}$ & $\langle c \rangle$ & $\frac{1}{g-1}$ & $\frac{1}{\ell_i-1}$ & $\langle c \rangle$\\
\midrule
$0.09$ && 5\%  & \textbf{100\%} & 5.77 & 3\%  & \textbf{100\%} & 6.28 & 6\%  & \textbf{100\%} & 5.96\\
$0.13$ && 26\%  & \textbf{100\%} & 4.48 & 30\%  & \textbf{100\%} & 4.18 & 37\%  & \textbf{100\%} & 3.50\\
$0.17$ && 50\%  & \textbf{100\%} & 3.39 & 62\%  & \textbf{100\%} & 2.60 & 69\%  & \textbf{100\%} & 2.15\\
$0.21$ && 67\%  & \textbf{100\%} & 2.61 & 78\%  & \textbf{100\%} & 1.86 & 89\%  & \textbf{100\%} & 1.51\\
$0.25$ && 83\%  & \textbf{100\%} & 1.97 & 90\%  & \textbf{100\%} & 1.45 & 98\%  & \textbf{100\%} & 1.22\\
$0.29$ && 85\%  & \textbf{100\%} & 1.63 & 94\%  & \textbf{100\%} & 1.25 & 97\%  & \textbf{100\%} & 1.12\\
$0.33$ && 95\%  & \textbf{100\%} & 1.30 & 99\%  & \textbf{100\%} & 1.11 & \textbf{100\%}  & \textbf{100\%} & 1.04\\
$0.37$ && 98\%  & \textbf{100\%} & 1.15 & \textbf{100\%} & \textbf{100\%} & 1.07 & \textbf{100\%} & \textbf{100\%} & 1.03\\
$0.41$ && 99\%  & \textbf{100\%} & 1.08 & \textbf{100\%} & \textbf{100\%} & 1.03 & \textbf{100\%} & \textbf{100\%} & 1.01\\
$0.45$ && \textbf{100\%}  & \textbf{100\%} & 1.05 & \textbf{100\%} & \textbf{100\%} & 1.01 & \textbf{100\%} & \textbf{100\%} & 1.00\\
\midrule
\end{tabular*}
\label{tab:geo}
\end{table}

\section{Discussion}\label{sec12}

The proposed mathematical programs show the versatility with which the GBP can be formulated. Each program highlights different aspects of the problem, and their relative advantages depend on the context in which they are applied. For example, PROP-MILP is intuitive and reflects the propagation process, but at the expense of requiring $2Un$ variables, which makes it less practical for modeling large graphs. In contrast, COV-CSP and COV-ILP are more compact formulations that leverage the relationship between the GBP and the CMCP. While COV-CSP is infeasible when the burning number is underestimated, COV-ILP is always feasible. A more practical program is GBP-ILP, which avoids using binary search and maintains a low number of variables, $Un$. From a classical computing point of view, GBP-ILP is ``better'' than all other programs. However, that does not undermine the potential of all other programs. For instance, the introduced QUBO problems arise directly from COV-CSP. Table \ref{tab:discussion} reports the main attributes of the proposed mathematical programs.

\begin{table}[h]
\caption{Main attributes of the proposed mathematical programs}\label{tab:discussion}
\begin{tabular*}{\textwidth}{@{\extracolsep\fill}ccccc}
\toprule%
program & variables & constraints & disadvantages & advantages \\
\midrule
\multirow{3}{*}{PROP-MILP} & \multirow{3}{*}{$2Un$} & \multirow{3}{*}{$Un+U+n$} & \multirow{3}{3.5cm}{\centering It has more variables and constraints than the other programs} & \multirow{3}{3.5cm}{\centering It is conceptually simple and explicitly codifies the propagation process} \\
&&&&\\
&&&&\\
&&&&\\
\multirow{4}{*}{COV-CSP} & \multirow{4}{*}{$gn$} & \multirow{4}{*}{$g+n$} & \multirow{4}{3.5cm}{\centering If $g<b(G)$, the program is infeasible. Besides, it requires being embedded into a binary search} & \multirow{4}{3.5cm}{\centering It is the simpler program} \\
&&&&\\
&&&&\\
&&&&\\
&&&&\\
\multirow{3}{*}{COV-ILP} & \multirow{3}{*}{$gn$} & \multirow{3}{*}{$g+n-1$} & \multirow{3}{3.5cm}{\centering It requires being embedded into a binary search} & \multirow{3}{3.5cm}{\centering Regardless of the value of $g$, it is always feasible} \\
&&&&\\
&&&&\\
&&&&\\
\multirow{3}{*}{GBP-ILP} & \multirow{3}{*}{$Un$} & \multirow{3}{*}{$2U+n-1$} & \multirow{3}{3.5cm}{\centering -} & \multirow{3}{3.5cm}{\centering It does not need being embedded into a binary search} \\
&&&&\\
&&&&\\
&&&&\\
\multirow{4}{*}{sQUBO} & \multirow{4}{*}{$(g + \log_2 g)n$} & \multirow{4}{*}{-} & \multirow{4}{3.5cm}{\centering Requires $n\log_2 g$ extra slack variables. Besides, it requires being embedded into a binary search} & \multirow{4}{3.5cm}{\centering Unlike uQUBO, it does not need any fine tuning} \\
&&&&\\
&&&&\\
&&&&\\
&&&&\\
\multirow{4}{*}{uQUBO} & \multirow{4}{*}{$gn$} & \multirow{4}{*}{-} & \multirow{4}{3.5cm}{\centering Requires a fine tuning of its penalization terms. Besides, it requires being embedded into a binary search} & \multirow{4}{3.5cm}{\centering Unlike sQUBO, it does not need extra slack variables} \\
&&&&\\
&&&&\\
&&&&\\
&&&&\\
\midrule
\end{tabular*}
\end{table}

The proposed QUBO problems, sQUBO and uQUBO, extend the scope of GBP to quantum computing. sQUBO guarantees correctness at the expense of additional slack variables, whereas uQUBO significantly reduces the search space by removing them. However, this simplification introduces the risk of obtaining optimal solutions that are not burning sequences. Nevertheless, our experiments indicate that with a simple adjustment of penalization parameters, uQUBO achieves consistent optimality on Erdős–Rényi and geometric random graphs. This suggests that QUBO formulations may provide a practical avenue for solving the GBP on both classical and quantum hardware.

Finally, in Appendix~\ref{secA1} we show that Gurobi can solve challenging instances of GBP-ILP using a row generation method in reasonable amounts of time. For example, it solved a graph with more than 4 million vertices in no more than 19 minutes. More challenging instances, such as large square grids, were solved in up to 2.6 hours.

\section{Conclusion}\label{sec13}

This paper presents an overview of mathematical programs for the GBP. Compared to existing programs in the literature, the proposed ones are conceptually simpler and require fewer variables, making them attractive for both theoretical analysis and practical computation.

The relevance of these programs lies not only in their immediate applicability but also in their potential to inspire methods for related contagion and propagation models. In addition, variants of the GBP can be modeled similarly. However, limitations remain: large-scale instances continue to challenge exact methods, and the tuning of penalization parameters in QUBO models with unbalanced penalization still requires heuristic guidance. Future research should explore hybrid classical–quantum approaches, as well as heuristic, metaheuristic, and matheuristic search. As a by-product, we corroborated the burning number of some large graphs reported by Pereira et al.~\cite{pereira_et_al:LIPIcs.ESA.2024.94} and found the burning number of some other challenging graphs.

\paragraph{Funding}
This research was funded by the Secretaría de Ciencia, Humanidades, Tecnología e Innovación (SECIHTI); grant number 914787.

\paragraph{Conflict of interest}
The authors have no competing interests to declare that are relevant to the content of this article.

\paragraph{Data availability}
The data can be consulted at \url{https://networkrepository.com/} and \url{https://github.com/NodesOnFire/GBP-ILP}

\paragraph{Code availability}
The implemented code can be consulted at \url{https://github.com/NodesOnFire/GBP-ILP}

\paragraph{Acknowledgements}
The authors acknowledge Secretaría de Ciencia, Humanidades, Tecnología e Innovación (SECIHTI) and Instituto Nacional de Astrofísica, Óptica y Electrónica (INAOE) for providing the necessary resources for the development of this research.

\begin{appendices}

\section{Solving some challenging instances}\label{secA1}

This section reports the time required by Gurobi to solve GBP-ILP (see Section~\ref{sec:ilp2}) using a row generation technique over some challenging graphs from the literature~\cite{nr,pereira_et_al:LIPIcs.ESA.2024.94}. These graphs are divided into two categories: large social networks and square grids. Table~\ref{tab:networks} shows some of the main attributes of these graphs, where $n$ is the number of vertices, $m$ is the number of edges, $\rho$ is the density, $\langle k \rangle$ is the average degree, $\langle C \rangle$ is the average clustering coefficient, and $b(G)$ is the burning number. Table~\ref{tab:setup} shows the experimental setup. Gurobi was executed with its default parameters, except for the LazyConstraints flag, which was set to 1. The code can be consulted at \url{https://github.com/NodesOnFire/GBP-ILP}.

\begin{table}[h]
\caption{Main attributes of some large social networks and square grids}\label{tab:networks}
\begin{tabular*}{\textwidth}{@{\extracolsep\fill}ccccccc}
\toprule%
\multicolumn{7}{@{}c@{}}{Large social networks}\\
\midrule
name&$n$&$m$&$\rho$&$\langle k \rangle$&$\langle C \rangle$&$b(G)$\\
\midrule
soc-youtube-snap&1134890&2987624&4.639E-6&5.265&0.172&11\\
soc-pokec&1632803&22301964&1.673E-5&27.317&0.122&8\\
socfb-B-anon&2937612&20959854&4.858E-6&14.270&0.209&8\\
socfb-A-anon&3097165&23667394&4.935E-6&15.283&0.209&7\\
soc-livejournal&4033137&27933062&3.434E-6&13.852&0.327&13\\
\midrule
\multicolumn{7}{@{}c@{}}{Square grids}\\
\midrule
name&$n$&$m$&$\rho$&$\langle k \rangle$&$\langle C \rangle$&$b(G)$\\
\midrule
grid\_50x50&2500&4900&1.569E-3&3.920&0&17\\
grid\_60x60&3600&7080&1.093E-3&3.933&0&19\\
grid\_70x70&4900&9660 &8.048E-4&3.943&0&21\\
grid\_80x80&6400&12640&6.173E-5&3.950&0&23\\
grid\_90x90&8100&16020&4.884E-4&3.956&0&25\\
\midrule
\end{tabular*}
\end{table}

\begin{table}[h]
\centering
\caption{Experimental setup}\label{tab:setup}
\begin{tabular*}{0.5\textwidth}{@{\extracolsep\fill}cc}
\toprule%
Operating system & Ubuntu 25.04\\
Processor & intel i9-12900K\\
RAM & DDR5 128 GB\\
Language & C++\\
Compiler & GNU GCC 14.2.0\\
Solver & Gurobi 12.0.3\\
\midrule
\end{tabular*}
\end{table}

Based on the work of Pereira et al.~\cite{pereira_et_al:LIPIcs.ESA.2024.94}, we used Gurobi to solve GBP-ILP with a row generation technique that adds coverage constraints (see Constraints (\ref{MP:4:3})) on demand. This is achieved by loading the program into Gurobi with a few initial coverage constraints and then adding some of the violated constraints as required. Actually, we added both violated and unviolated constraints to keep a more representative set of vertices. The intuition behind this idea is that a few vertices are often representative of the whole. Therefore, by burning them, many of the remaining vertices are burned too. As a rule of thumb, we start by loading $2U$ coverage constraints; each corresponding to each vertex in a greedy permutation of length $2U$ starting with vertex $v_1$, where a greedy permutation is a list of vertices such that each vertex is the farthest one from the previous vertices~\cite{sheehy2020one,gonzalez1985clustering,dyer1985simple}. Each time a new incumbent solution is found, $U$ additional constraints are introduced to the model. As before, these correspond to a greedy permutation beginning with any vertex tied to a violated constraint. Table~\ref{tab:experimentalresults} shows the time and number of constraints required by Gurobi to return an optimal burning sequence; a dash symbol (\textendash) indicates that the experiment was terminated by the operating system due to an insufficient memory exception. 

\begin{table}[h]
\caption{Execution time ($t$) and number of coverage constraints (\#cc) required by Gurobi for solving GBP-ILP with row generation}\label{tab:experimentalresults}
\begin{tabular*}{\textwidth}{@{\extracolsep\fill}ccccccccc}
\toprule%
\multirow{4}{*}{graph} & \multicolumn{8}{@{}c@{}}{$U$}\\
\cmidrule{2-9}
& \multicolumn{2}{@{}c@{}}{$b(G)$} & \multicolumn{2}{@{}c@{}}{$b(G)+1$} & \multicolumn{2}{@{}c@{}}{$b(G)+2$} & \multicolumn{2}{@{}c@{}}{$b(G)+3$}\\
\cmidrule{2-3}\cmidrule{4-5}\cmidrule{6-7}\cmidrule{8-9}
& $t$ & \#cc & $t$ & \#cc & $t$ & \#cc & $t$ & \#cc\\
\midrule
soc-youtube-snap&2m&22&3m&36&3m&26&23m&42\\
soc-pokec&3m&32&4m&36&18m&20&26m&22\\
socfb-B-anon&5m&32&7m&36&5m&40&13m&44\\
socfb-A-anon&4m&35&7m&64&10m&54&\textendash&\textendash\\
soc-livejournal&16m&39&15m&56&19m&45&\textendash&\textendash\\
\midrule
grid\_50x50&3m&1275&1m&936&2m&1178&7m&1460\\
grid\_60x60&14m&1938&4m&1360&24m&1554&18m&1716\\
grid\_70x70&13m&2121&10m&2090&1.2h&2921&14m&2448\\
grid\_80x80&39m&2461&1.1h&2712&48m&3100&1.3h&3718\\
grid\_90x90&2.3h&3975&1.9h&3666&2.6h&3456&1.6h&2912\\
\midrule
\end{tabular*}
\end{table}

As Table~\ref{tab:experimentalresults} shows, Gurobi is capable of solving challenging instances of the GBP-ILP in reasonable amounts of time by using a row generation technique. In summary, it solved the GBP on graphs with more than 4 million vertices in no more than 19 minutes and on grids with up to 8100 vertices in less than 2.6 hours. Although Table~\ref{tab:experimentalresults} raises many interesting questions, such as why highly structured graphs (like square grids) appear to be more difficult to solve, we defer those inquiries for future research.

\end{appendices}

\bibliographystyle{unsrt}  
\bibliography{sn-bibliography}

\end{document}